\documentclass{appolb}
\usepackage{graphicx}
\usepackage[toc,page]{appendix}
\usepackage[english]{babel}
\usepackage[utf8]{inputenc}
\usepackage{t1enc}
\usepackage{amsmath, amsthm, amssymb}
\usepackage{cite}
\usepackage{anysize}
\marginsize{2cm}{2cm}{2cm}{3.5cm} 
\usepackage{hyperref}

\graphicspath{{figures/}}

\begin{document}

\newcommand{\D}{\partial}
\newcommand{\dt}{\frac{\D}{\D t}}
\newcommand{\dxi}{\frac{\D}{\D x_i}}
\newcommand{\dx}{\frac{\D}{\D x}}
\newcommand{\dy}{\frac{\D}{\D y}}
\newcommand{\dz}{\frac{\D}{\D z}}
\newcommand{\rest}{\: \: \:}
\newcommand{\el}{\nonumber \\}
\newcommand{\fok}{^{\circ}}
\newcommand{\omg}{\omega_{\textmd{g}}}
\newcommand{\jnyil}{\rightarrow}

\title{HBT radii from multipole Buda-Lund model
\thanks{Presented  at  the  XI  Workshop  on  Particle  Correlations  and  Femtoscopy, 3-7 November 2015, Centre for Innovation and Technology Transfer Management, Warsaw University of Technology}
}
\author{S\'{a}ndor L\"{o}k\"{o}s$^1$, M\'{a}t\'{e} Csan\'{a}d$^1$, Boris Tom\'{a}\v{s}ik$^{2,3}$, Tam\'{a}s Cs\"{o}rg\H{o}$^{4,5}$
\address{$^1$E\"{o}tv\"{o}s Lor\'{a}nd University, Budapest, Hungary\\
$^2$Univerzita Mateja Bela, Bansk\'a Bystrica, Slovakia \\
$^3$FNSPE, Czech Technical University, Prague, Czech Republic \\
$^4$Wigner RCP of the HAS, Budapest, Hungary \\
$^5$K\'aroly R\'obert College, Gy\"ongy\"os, Hungary}
}
\pagestyle{plain}
\maketitle
\begin{abstract}
The Buda-Lund model describes an expanding hydrodynamical system with ellipsoidal symmetry and fits the observed elliptic flow and oscillating HBT radii successfully. The ellipsoidal symmetry can be characterized by the second order harmonics of the transverse momentum distribution, and it can be also observed in the azimuthal oscillation of the HBT radii measured versus the second order reaction plane. The model may have to be changed to describe the experimentally indicated higher order asymmetries. In this paper we detail an extension of the Buda Lund hydro model to investigate higher order flow harmonics and the triangular dependence of the azimuthally sensitive HBT radii.
\end{abstract}
  
\section{Introduction}

The investigation of the correlation functions is a useful tool to measure properties of the strongly interacting quark-gluon plasma (sQGP) such as the size or the asymmetries of the medium. The method originally had been discovered by Hanbury-Brown and Twiss \cite{HBT} in radioastronomy and Goldhaber et. al. developed it to measure the size of the source in heavy ion reactions \cite{PhysRev.120.300}. If the system is not spherically symmetric there could be azimuthal dependence of the HBT radii. In the framework of the Buda-Lund model the $2^{nd}$ order oscillation of these radii can be investigated (see \cite{Csanad:2008af}) but the experiments show that (e.g. in \cite{Adare:2014vax}) there are triangular ($3^{rd}$ order) oscillations too. To observed the azimuthal dependence of the radii it has to be investigated in the corresponding reaction plane. In the extended Buda-Lund model $v_3$ has non-zero value contrary to the elliptical one and the third order oscillation amplitudes can be investigated too.

\section{Description of the geometry}

\begin{figure}[htb]
\centerline{%
\includegraphics[width=7cm]{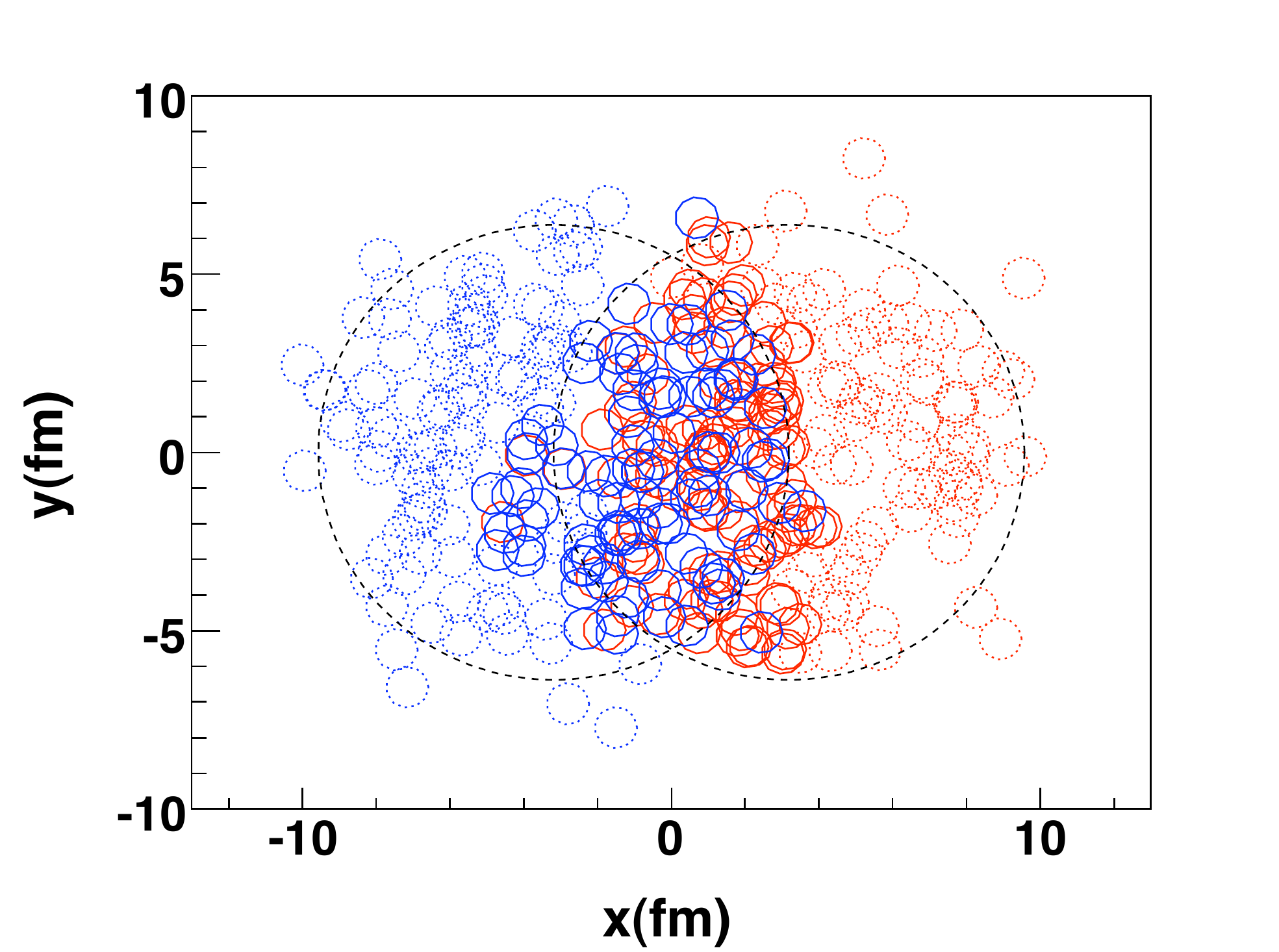}
\includegraphics[width=5cm]{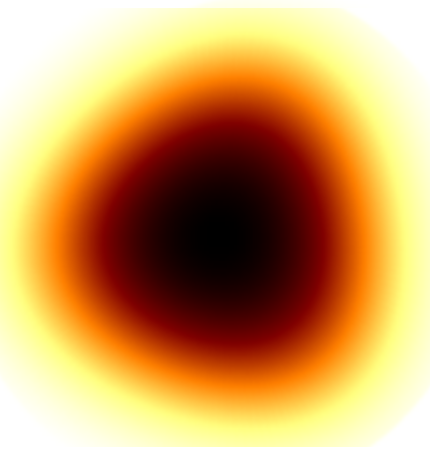}
\includegraphics[width=5cm]{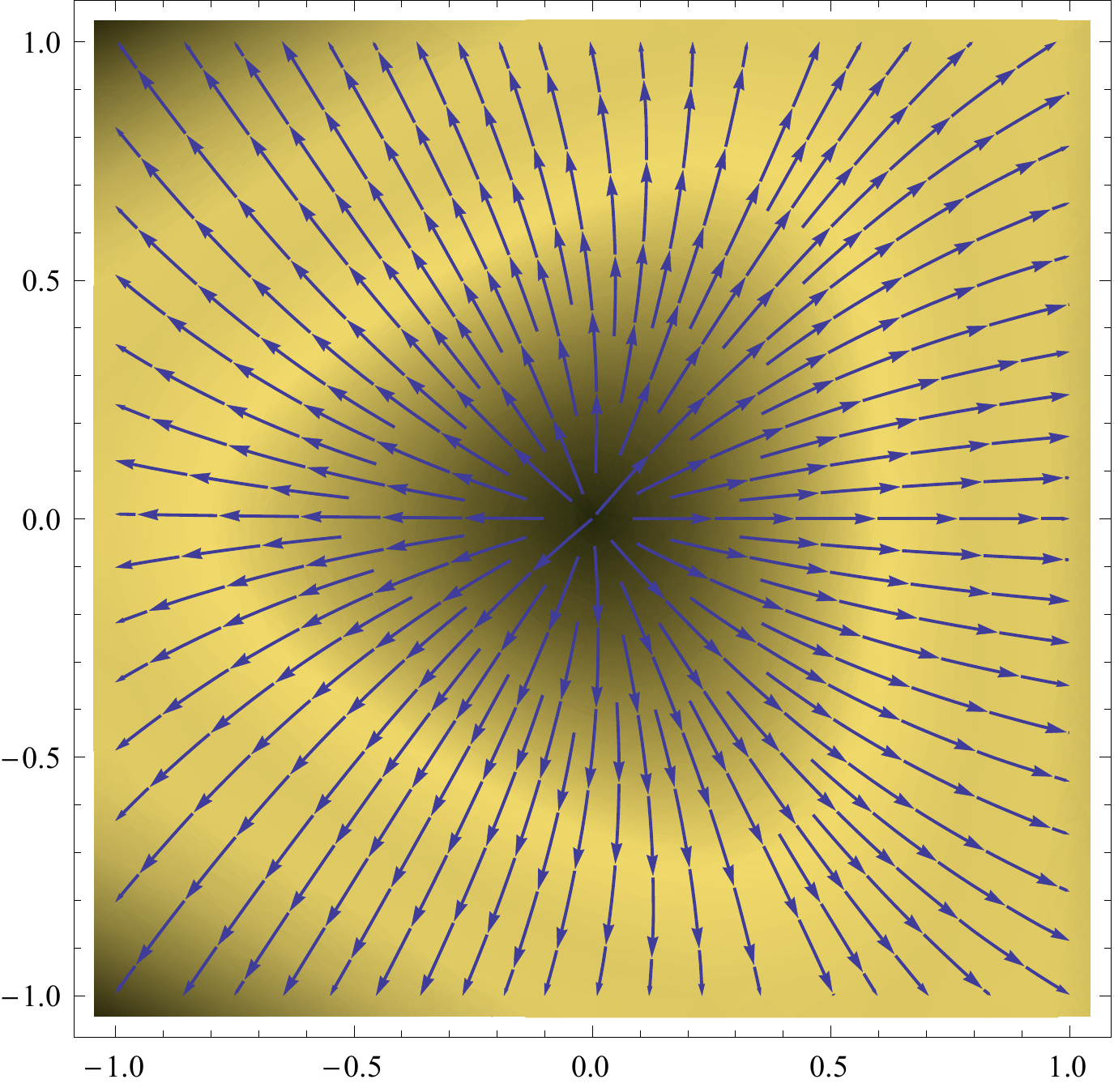}}
\caption{Plot of a Glauber simulation in $Au+Au$ at RHIC energy (Ref. \cite{Loizides:2014vua}). The triangular density profile (Ref. \cite{Csanad:2014dpa}) is in the center figure where the value of the asymmetry parameter is $\epsilon_3=0.5$. On the right hand side there is the triangular flow pattern with $\chi_3=0.2$}
\label{fig:geometry}
\end{figure}

Generally the source function assumed in hydrodynamical model is
\begin{align}
S(x,p)d^4x=\frac{g}{(2\pi)^3}\frac{p^\mu d^4\Sigma_\mu(x)}{B(x,p)+s_q} \: \: \: \: \textmd{with } \: \: \: \: B(x,p)=\exp{\left[\frac{p_\mu u^\mu(x)-\mu}{T(x)}\right]}
\label{eq:source}
\end{align}
where $g$ is the degeneracy factor, $p^\mu d^4\Sigma_\mu(x)$ is the Cooper-Frye factor, $B(x,p)$ is the Boltzmann-distribution and $s_q$ is the usual quantum statistic term. In the hydrodynamical investigation of the sQGP there are two kinds of asymmetries that can be important to understand the phenomenology of the dynamics, the asymmetry of the space-time distribution and the velocity field distribution. The asymmetry of the space-time distribution is ensured by a scale parameter $s$. An example spatial distribution is shown in Fig. \ref{fig:geometry}. We can generalize the model through this parameter.
\\
\linebreak
Let us write up $s$ in a cylindrical coordinate system instead of Cartesian with $r^2=r_x^2+r_y^2$ and $\cos(\varphi)=\frac{r_x}{r}$, and prescribe the azimuthal angle dependence. A general case can be written up like:
\begin{align}
s=\frac{r^2}{R^2}\left( 1+ \sum_n \epsilon_n \cos(n(\varphi-\Phi_n)) \right) + \frac{r_z^2}{Z^2}
\end{align}
The ellipsoidal case corresponds to $n=2$. There is a known solution of the hydrodynamics with similar generalization of the $s$ in Ref. \cite{Csanad:2014dpa} however in that solution the velocity field distribution was not generalized to arbitrary asymmetry. We can generalize the velocity field in the present framework with the generalization of the velocity field potential $\Phi$, defined as $u_\mu = \gamma(1, \nabla \Phi)$. A general case can be expressed:
\begin{align}
\Phi=Hr^2 \left( 1 + \sum_n \chi_n\cos(n(\varphi-\Phi_n)) \right) + H_z r_z^2
\end{align}

\section{Observables}

\begin{figure}[htb]
\centerline{%
\includegraphics[width=16cm]{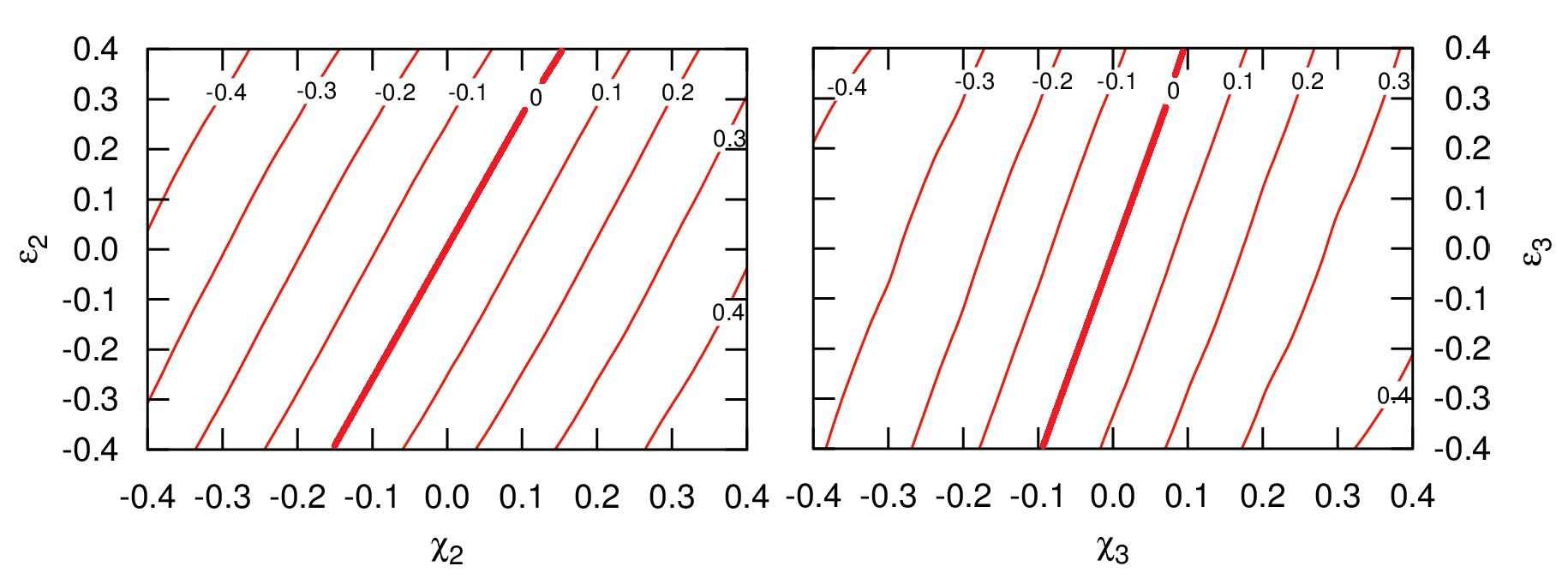}}
\caption{Elliptic flow $v_2$ at $p_t=1000$ MeV$/$c as function of $\chi_2$ and $\epsilon_2$ (left). Triangular flow $v_3$ at $p_t=1000$ MeV$/$c as function of $\chi_3$ and $\epsilon_3$ (right)}
\label{Fig:v2_v3_cont}
\end{figure}

The invariant transverse momentum distribution, the $n$-order flows and the azimuthally sensitive HBT radii can be derived from the model. The asymmetries and their effects on the observables can be investigated versus the proper reaction plane. If we explore quantities respect to the $2^{nd}$ order reaction plane we have to average on the angle between the $2^{nd}$ and $3^{rd}$ order reaction plane and vice versa.
\\
The flow coefficients can be obtained from the Fourier-series of the invariant momentum distribution and can be expressed in the form $v_n(p_t)=\langle \cos(n\alpha) \rangle_{N_1(p)}$. The investigations show that the $n$-th order asymmetry coefficients affect only the $n$-th order flow, however both of the spatial and the velocity field asymmetry have effects. For example $\epsilon_2$ and $\chi_2$ has effect on the $v_2$ but has no effect on $v_3$. This phenomenon is illustrated in Fig. \ref{Fig:v2_v3_cont}. There is an entanglement of the spatial and the velocity field asymmetry so the values of these cannot be extracted only from the measurements of the flows.
\\
The azimuthally sensitive HBT radii are important in the survey of the geometry and the size of the source. Generally the two particle momentum correlation function can be expressed as the Fourier-transform of the source function. Based on Ref. \cite{Plumberg:2013nga} the HBT radii can be calculated as
\begin{align}
R_{out}^2 = \langle r_{out}^2 \rangle - \langle r_{out} \rangle^2 \: \: \: \textmd{and} \: \: \: R_{side}^2 = \langle r_{side}^2 \rangle - \langle r_{side} \rangle^2
\label{eq:hbt_radii_av}
\end{align}
where $r_{out}=r\cos(\varphi-\alpha)-\beta_t t$ and $r_{side}=r\sin(\varphi-\alpha)$, $\varphi$ is the spatial angle and $\alpha$ is the momentum angle. Average $\langle \cdot \rangle$ means for example $\langle r \rangle = \int rS(x,p)d^4x$ where in our case $S(x,p)$ is defined in Eq. \eqref{eq:source}. We can calculate the azimuthal angle dependence of the HBT radii from Eq.  \eqref{eq:hbt_radii_av} and parametrize them with the following functions:
\begin{align}
\textmd{Elliptical case: } R_{out}^2 &=R_{o,0}^2+R_{o,2}^2\cos(2\alpha)+R_{o,4}^2\cos(4\alpha)+R_{o,6}^2\cos(6\alpha) \\
\textmd{Triangular case: } R_{out}^2 &=R_{o,0}^2+R_{o,3}^2\cos(3\alpha)+R_{o,6}^2\cos(6\alpha)+R_{o,9}^2\cos(9\alpha)
\end{align}
The additional terms are caused by the rotation to the $(out,side,long)$ system as it detailed in Ref. \cite{Tomasik:2002rx}. These amplitudes are affected by both of the asymmetry parameters so there is also a mixing in this case. Yielding the value of the parameters from only the azimuthally sensitive HBT radii measurements is not possible but combined with the flow measurements the value of the asymmetry coefficients can be determined.

\section{Summary and acknowledgements}

We can generalized the space-time and velocity field distribution in the framework of the Buda-Lund hydro model. Observables can be derived from the extended model and the effects of the higher order asymmetries can be investigated on the flows and the azimuthally sensitive HBT radii. However there is a mixing between the different type of the asymmetry coefficients but combining the measurements together could be a method to disentangle the value of these parameters. This work was partially supported by the OTKA NK 101438 grant.

\begin{figure}[htb]
\centerline{%
\includegraphics[width=16cm]{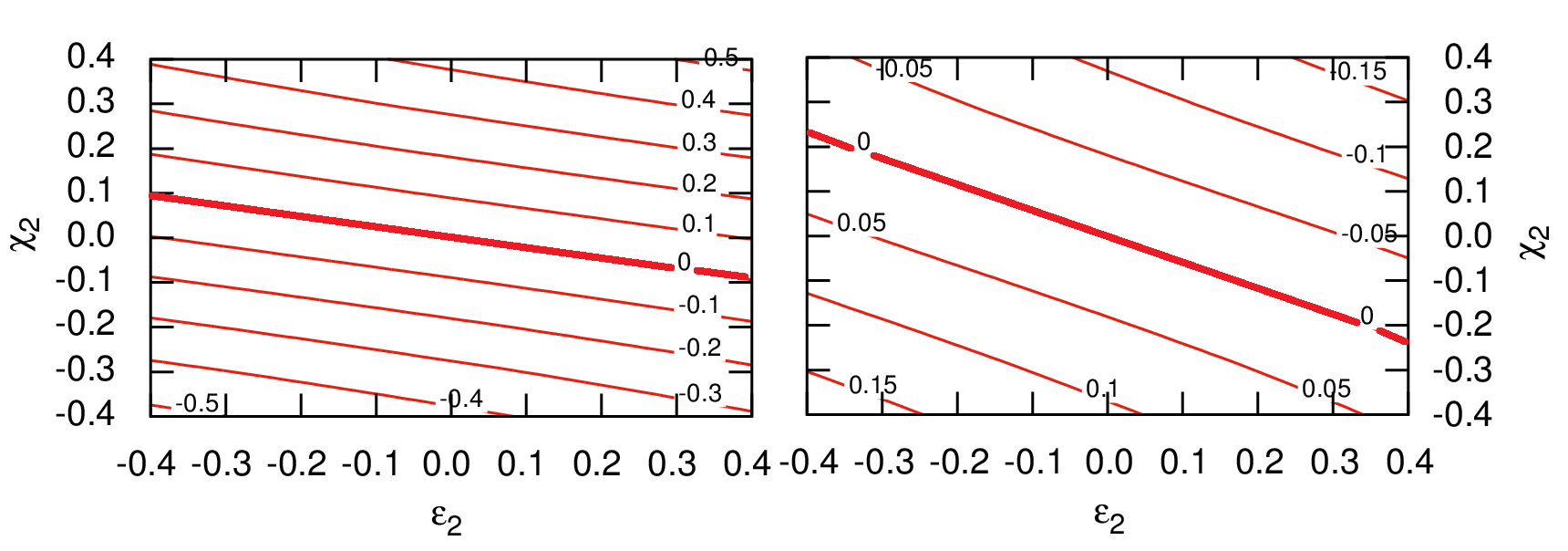}}
\caption{The parameter dependence of the second order oscillations of the $R^2_{out,2}/R^2_{out,0}$ (left) and the $R^2_{side,2}/R^2_{side,0}$ (right) at $p_t=300$ MeV$/$c.}
\label{Fig:azim_HBT_2}
\end{figure}

\begin{figure}[htb]
\centerline{%
\includegraphics[width=16cm]{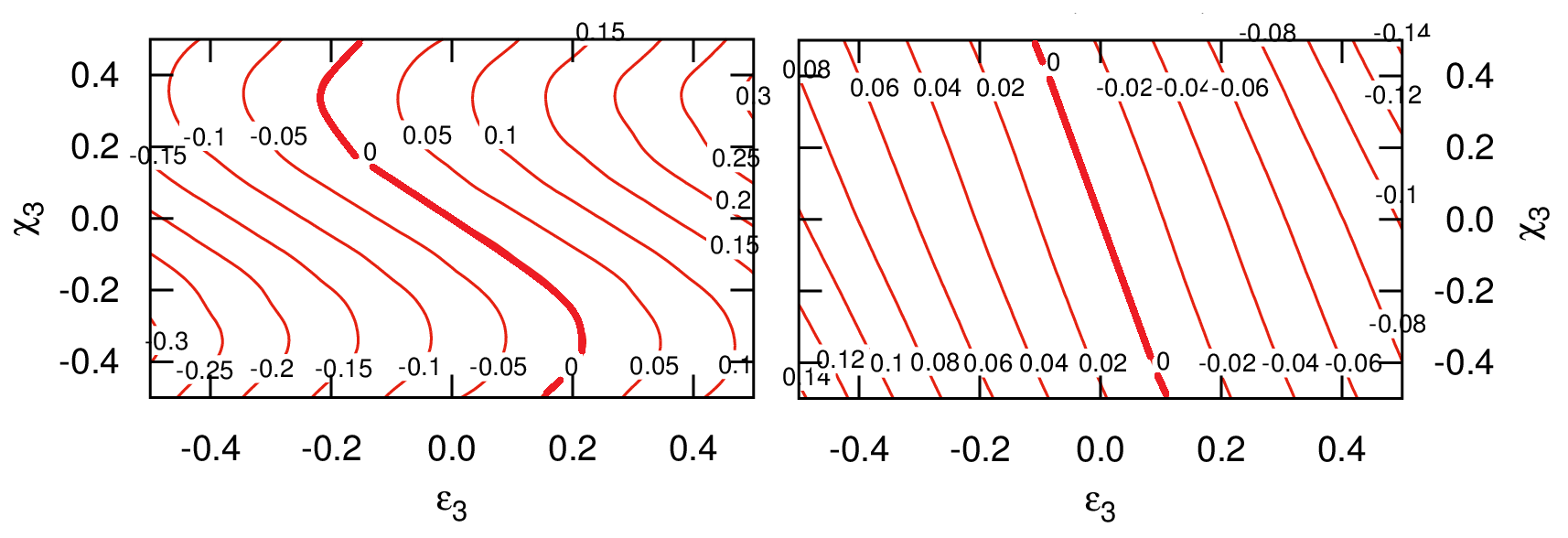}}
\caption{The parameter dependence of the third order oscillations of the $R^2_{out,3}/R^2_{out,0}$ (left) and the $R^2_{side,3}/R^2_{side,0}$ (right) at $p_t=300$ MeV$/$c.}
\label{Fig:azim_HBT_3}
\end{figure}

\end{document}